\documentclass{article}

\usepackage{arxiv}

\usepackage[utf8]{inputenc} % allow utf-8 input
\usepackage[T1]{fontenc}    % use 8-bit T1 fonts
\usepackage{hyperref}       % hyperlinks
\usepackage{url}            % simple URL typesetting
\usepackage{booktabs}       % professional-quality tables
\usepackage{amsfonts}       % blackboard math symbols
\usepackage{nicefrac}       % compact symbols for 1/2, etc.
\usepackage{microtype}      % microtypography
\usepackage{lipsum}		% Can be removed after putting your text content
\usepackage{graphicx}
\usepackage{natbib}
\usepackage{doi}
\usepackage[most]{tcolorbox}

\definecolor{lightgray}{RGB}{217,217,217}
\definecolor{lightlightgray}{RGB}{235,235,235}

\newtcolorbox[auto counter]{finding}[1][]{
  enhanced,
  breakable,
  left=4pt,right=4pt,top=6pt,bottom=6pt,colback=gray!5,colframe=gray!25,  arc=1mm, 
  #1
}

\title{Log severity level classification: 
an approach for systems in production}

\author{ \href{https://orcid.org/0000-0001-8115-9512}{\includegraphics[scale=0.06]{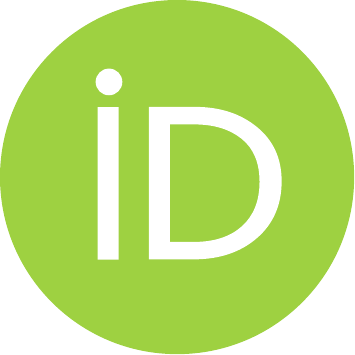}\hspace{1mm}Eduardo Mendes}\\
	Départment d'informatique et mathemátique\\
	Université du Québec à Chicoutimi\\
	Saguenay, QC \\
	\texttt{eduardo.mendes-de-oliveira1@uqac.ca} \\
	\And
	\href{https://orcid.org/0000-0002-8355-1494}{\includegraphics[scale=0.06]{orcid.pdf}\hspace{1mm}Fabio Petrillo} \\
	Départment d'informatique et mathemátique\\
	Université du Québec à Chicoutimi\\
	Saguenay, QC \\
	\texttt{fabio@petrillo.com} \\
}

\renewcommand{\shorttitle}{Log severity level classification: an approach for systems in production}

\hypersetup{
pdftitle={Log severity level classification: an approach for systems in production},
pdfsubject={cs.SE},
pdfauthor={Eduardo ~Mendes, Fabio ~Petrillo},
pdfkeywords={log severity level, log classification, log entry, log statement, logging},
}

\begin{document}
\maketitle

\begin{abstract}
% Context:
\textbf{Context: }Logs are often the primary source of information for system developers and operations engineers to understand and diagnose the behavior of a software system in production. In many cases, logs are the only evidence available for fault investigation.
% Problem:
\textbf{Problem:} However, the inappropriate choice of log severity level can impact the amount of log data generated and, consequently, quality.  This storage overhead can impact the performance of log-based monitoring systems, as excess log data comes with increased aggregate noise, making it challenging to utilize what is actually important when trying to do diagnostics.
% Objective:
\textbf{Goal:} This research aims to decrease the overheads of monitoring systems by processing the severity level of log data from systems in production.
% Approach
\textbf{Approach:} To achieve this goal, we intend to deepen the knowledge about the log severity levels and develop an automated approach to log severity level classification, demonstrating that reducing log severity level “noise” improves the monitoring of systems in production. 
% Impact:
\textbf{Conclusion:} We hope that the set of contributions from this work can improve the monitoring activities of software systems and contribute to the creation of knowledge that improves logging practices.
\end{abstract}

% keywords can be removed
\keywords{log severity level \and log classification \and log entry \and log statement \and logging }

\section{Introduction}

Logs are often the primary source of information for system developers and operators to understand and diagnose the behavior of a software system \citep{IST/EL2020/systematic}. By examining logs, developers can identify bugs more quickly \citep{ICRACS/BUSHONG-2020/matching}. For operations engineers, in many cases, logs are the only evidence to investigate the system's failure and understand a system's run-time behavior \citep{TOCS/YUAN-2012/diagnosability, EMSE/YAO-2020/compressors}. According to \cite{ICSE/LIN2016/log-clustering}, \textit{ “engineers need to examine the  recorded  logs  to  gain  insight  into  the  failure,  identify  the problems,  and  perform  troubleshooting”}. 

\begin{figure}[h]
\centerline{\includegraphics[width=1\linewidth]{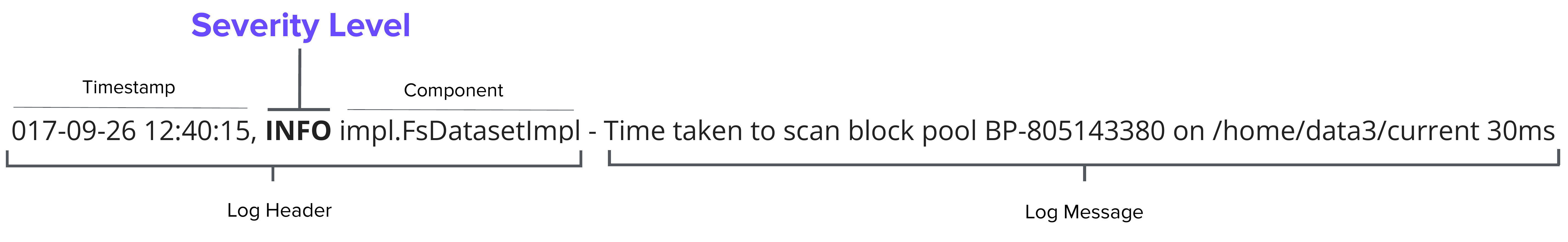}}
\caption[Log severity level in a log entry.]{Log severity level in a log entry. Adapted from \citet{IST/EL2020/systematic}.} 
\label{IMAGE/severity-level-in-a-log-entry}
\end{figure}

As presented in Figure \ref{IMAGE/severity-level-in-a-log-entry}, each log entry is usually composed of time-stamp, severity level, software component, and log message. \textbf{\textit{Severity levels}} indicate the degree of severity of the log message \citep{SPE/KIM2020/automatic}. For example, a less severe level is used to indicate that the system behaves as expected, while a more severe level is used to indicate that a problem has occurred \citep{ICSE/CHEN2017/characterizing-antipatterns}. Among the causes of generating less or more evidence, less or more log entries, we can observe the choice of log severity level.

\subsection{Problem Statement and Research Motivation}

\begin{figure}[h]
\centerline{\includegraphics[width=1\linewidth]{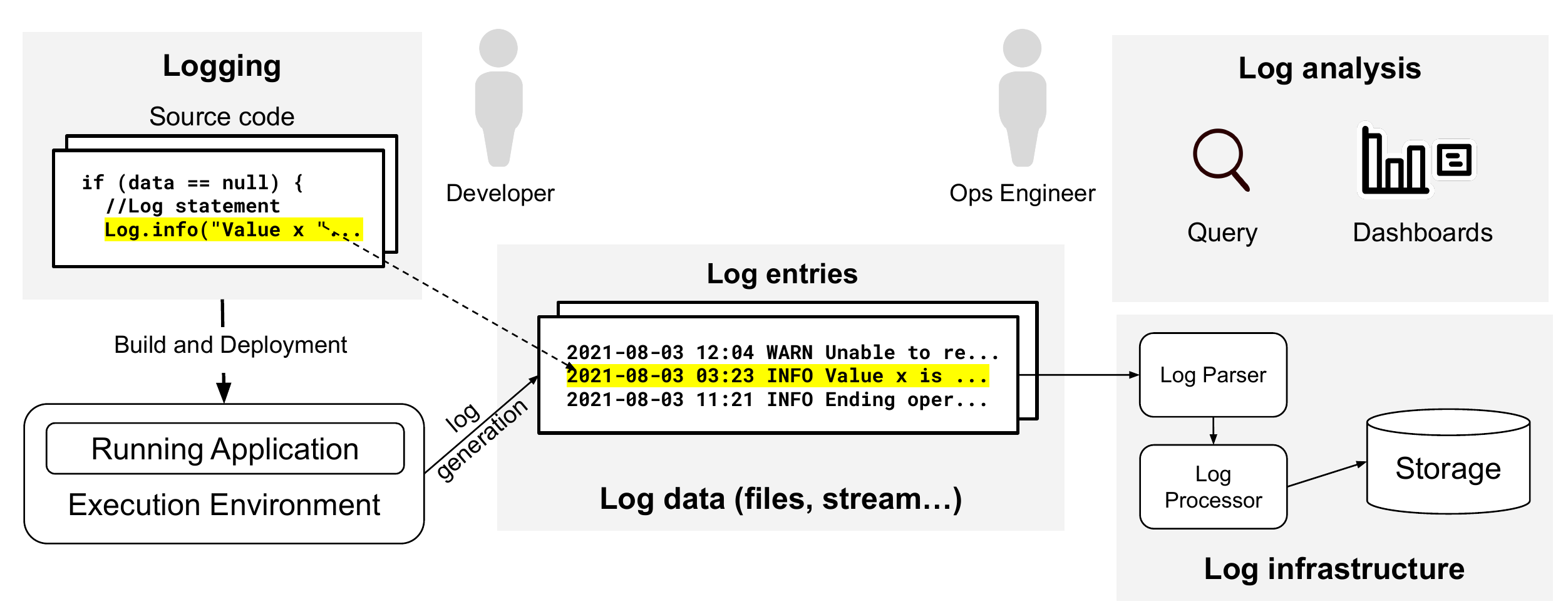}}
\caption[Overview of the life-cycle of log data.]{Overview of the life-cycle of log data, based on \citet{PEERJ/CANDIDO-2021/LogBasedMonitoringMapping}.} 
\label{IMAGE/log-data-lifecycle}
\end{figure}

% Log entries record variable states and events that occur during a software system's execution \citep{ARXIV/MENDES-2021-log-levels-matter}. 
As presented on \textbf{Figure \ref{IMAGE/log-data-lifecycle}}, the process to generate the log entries begins in the software development phase when developers choose the points at which the \textit{log statements} \footnote{We use the term \textit{log statement} to identify the code commands that generate the log data. Typically, developers use \textit{logging libraries} to add log statements.} will transverse the source code \citep{EMSE-LI2018-studying}. A log entry will be generated for each chosen line code with a log statement when a software system is running, \textit{i.e.}, an entry will be added to the system \textit{log data} \footnote{We use log data to refer to a collection of log entries generated by a software system,
whether in log files, data streams, or other types of storage.} every time the execution reaches a log statement. These data can be parsed, processed and stored to be consumed in monitoring activities in systems under development or production, whether by systems or humans. Operations engineers use them to monitor, for example, whether the system is working or not, to analyze whether it is on the verge of failing, to identify behavior anomalies, to understand particularities during its operation through these data, \textit{i.e.}, to understand how the system behaves in  \citep{PEERJ/CANDIDO-2021/LogBasedMonitoringMapping}.

% Problem
The choice of severity level impacts the amount of log data that a software system produces
\citep{ICSE/LIN2016/log-clustering, ICSE/CHEN2017/characterizing-antipatterns, ESE/CHOWDHURY2018/exploratory, ESE/ZENG2019/studying}. 
For example, if a system is set to \textit{Warn} level, only statements marked with \textit{Warn} and higher levels (e.g., \textit{Error}, \textit{Fatal}) will be output \citep{ICSE/CHEN2017/characterizing-antipatterns}.

% The verbosity of the log chosen in the data generation process determines the amount of log information that the system will produce, whether it will produce information concerning all types of log levels or only a subset of the levels used \citep{ESE/CHEN2017/characterizing-logging-practices}.
% In addition to the severity level, another factor that can contribute to the generation of more log data is the lack of understanding about choosing the appropriate log level for a log message, for example, a genuine DEBUG message classified as an INFO message. 
% If this misunderstanding is recurrent in development, we can envision an important impact on the size of the generated artifacts. 

Developers spend significant time adjusting log severity levels \citep{ESE/KABINNA2018/examining}. After an initial choice, developers may modify the severity level re-evaluating how critical an event is \citep{ICSE/YUAN2012/characterizingLoggingPractices, SPSP/ZHAO2017/log20}. They can re-evaluate if a statement, initially classified as \textit{Info}, would actually be of \textit{Error} level, or if it would not be an intermediate level between the two levels, \textit{i.e.,} a \textit{Warn} \citep{SPSP/ZHAO2017/log20}. 
In this sense, when a developer choose severity levels inappropriately, the system can produce more log entries than it should, or the opposite, less log entries \citep{ESE/HASSANI2018/studying}. In both scenarios, the wrong choice of severity level can cause problems in the software system performance \citep{ICSE/CHEN2017/characterizing-antipatterns, ESE/LI2017/LogLevelChoose, ICSE/YUAN2012/characterizingLoggingPractices},
 and maintenance \citep{ESE/LI2017/LogLevelChoose, ASE/HE2018/characterizingNaturalLanguageDescriptions}.
Therefore, log files can be massive sizes, requiring large storage capacities from corporations \citep{CLOUD/YUAN2019/approach, ARXIV/GHOLAMIAN-2021/comprehensive}.

However, there is another consequence of the misunderstanding of the log severity levels: by generating more log data than necessary, the operations engineers and monitoring systems
% who use these files for analysis, monitoring purposes, or even the systems that use the log files as inputs
can potentially receive a high amount of noise \citep{ARXIV/MENDES-2021/towards-noisiness}, affecting log-based monitoring and diagnostics \citep{ESE/HASSANI2018/studying, ESE/LI2017/LogLevelChoose, ASWEC/RONG2018/logging}. According to \cite{USENIX/DING-2015/log2}, \textit{“(...) intensive logging could introduce a large amount of less ‘useful’ logs (i.e., the logs that are not useful for helping diagnose the performance issue under investigation)”} .

Among the factors that make choosing the severity level a challenge are: 
(i) lack of knowledge of how logs will be used \citep{COMACM/OLINER2012/advances}; 
(ii) lack of understanding how critical an event is \citep{ESE/ZENG2019/studying}; 
(iii) the ambiguity of certain events that seem to be related to multiple levels of severity \citep{ICSE/LIN2016/log-clustering, SPSP/ZHAO2017/log20}.
In addition, there is \textit{a lack of specifications and practical guidelines} for performing logging tasks in projects and industry \citep{ASE/HE2018/characterizingNaturalLanguageDescriptions, ASWEC/RONG2018/logging, ICSME/ANU2019/verbosityloglevels}. The consequence is that, in software development projects, \textit{“personal experience and preferences play an important role in logging practices”} \citep{ASWEC/RONG2018/logging}.

Although there are logging solutions such as Elasticsearch\footnote{https://www.elastic.co/elasticsearch/}, Logstash\footnote{https://www.elastic.co/logstash}, Kibana\footnote{https://www.elastic.co/kibana}, Grafana\footnote{(https://grafana.com} and FluentD\footnote{https://www.fluentd.org} which act as an infrastructure to process, visualize, and query logs, this whole scenario around what to log and where to log makes it challenging to monitor production systems and obtain information of the log data \citep{PEERJ/CANDIDO-2021/LogBasedMonitoringMapping}.

% Related works
Several studies propose solutions for the correct use of the log severity level. \citep{SPE/KIM2020/automatic} propose an approach to verify the appropriateness of the log severity levels. \cite{ESE/LI2017/LogLevelChoose}
propose a deep learning approach for log severity level prediction using the logging locations. \citep{ACMCASE/LI2020/where-shall} discussed where to apply logging locations and proposed a learning approach to provide code block logging suggestions. Other studies in the literature focus on \textit{“where to log”} such as \cite{SPSP/ZHAO2017/log20, ICSE/FU2014/developers} and \cite{TSE/LI2020/qualitative}. All these works propose approaches to the origin of this phenomenon, the code development phase. Furthermore, while developers generally rely on their own experience to insert log statements, existing approaches often use these statements as the oracle of their approaches \citep{ICSE/ZHU-2015/learning, ACMSAC/GHOLAMIAN2020/logging, ICSA/JIA-2018/smartlog}.
When we reflect on the systems already in production, we can infer that if incorrect log severity level classification were not addressed during development, then log data from systems in production will inherit this issue. 
  
\section{Proposal}
Considering the following aspects: (i) the lack of guidelines for creating of log statements, which consequently (ii) generate severity level noise in log data, and (iii) the impact of this on monitoring systems, our research addresses the following three hypothesis and goal:

\begin{tcolorbox}[colframe=gray!25, coltitle=black, arc=0mm, title=\textbf{Hypothesis 1}]
Processing the severity level of log entries in production improves the quality of log data.
\end{tcolorbox}

\begin{tcolorbox}[colframe=gray!25, coltitle=black, arc=0mm, title=\textbf{Hypothesis 2}]
    Log data with a lower noise level at log severity levels decrease overheads 
in log-based monitoring systems.
\end{tcolorbox}

\begin{tcolorbox}[colframe=gray!25, coltitle=black, arc=0mm, title=\textbf{Hypothesis 3}]
Dynamic log processing improves the diagnosis of monitoring systems.\end{tcolorbox}

\begin{tcolorbox}[colframe=gray!25, coltitle=black, arc=0mm, title=\textbf{Goal}]
    Improve software monitoring by processing log entries from systems in production.
\end{tcolorbox}

To this end, we propose an automatic approach to classify log severity levels for software systems in production.
According to \cite{IS/ROUDJANE-2021/detecting}, \textit{“processors are computational units that transform input streams into output streams.”} Our input is the log data, from which we intend to identify and process log entries in which the severity level is not adequate to its associated message. Our expected output is reclassified log data with with improved quality to be consumed by monitoring systems.

% We intend to identify and process log entries in which the severity level is not adequate to its associated message, decreasing the number of issues in the logs analyzed by monitoring systems (Figure \ref{fig:inputs-processor-monitoring}).

\begin{figure}[h]
 \centering
 \includegraphics[width=1\linewidth]{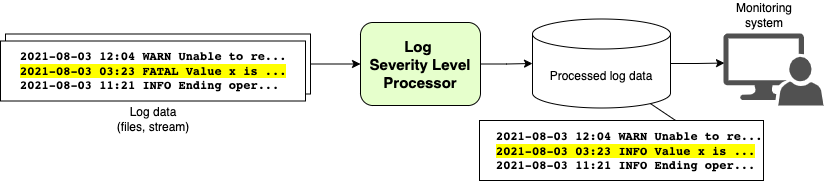}
 \caption{Log processor to classify severity levels}
 \label{fig:inputs-processor-monitoring}
\end{figure}

In order to find the heuristics and features that allow us to classify the log severity level of a log message
(i) we will further study what the log severity level is (logging libraries, peer-reviewed and grey literature),
(ii) we will explore public and industry log repositories, 
(iii) we will project and develop an approach to process and classify log severity levels in production (Figure \ref{fig:inputs-processor-monitoring}), and
(iv) we will report a complete validation of our approach.

\section{Conclusion}
Logs are essential assets for understanding development activities and monitoring the behavior of systems in production. However, there is still a lack in academia and industry for logging practices guidelines. As a consequence, log data can present issues such as severity level misclassification. Literature has shown interest in logs through different approaches, determining problems in coding log statements, automating the creation of log messages, suggesting severity levels. However, these approaches address the code development phase.

Our goal is to improve monitoring through automatic classification of log entries' severity levels generated by the systems. By reclassifying the misclassified entries, we hope to decrease the size of logs in production, for example, in situations where log entries are tagged with \textit{Info} level but are actually \textit{Debug} level. This would be a case where the processor would decrease the number of log entries provided to a monitoring system and, consequently, increase its performance. Another example would be when log entries are tagged with \textit{Fatal} level but are from a less severe level, such as \textit{Info}, which would cause red alerts in monitoring systems. Correct reclassification to \textit{Info} would contribute to monitoring quality and the analysis results.

We also envision that an automated approach can be beneficial in other scenarios. The first scenario would be to use it on log entries that do not have severity levels. A second example would be to establish code quality measures from the processing and analyze the severity level. The number of log entries with severity levels misclassified in the development phase can give us clues about the quality of the code. A third scenario would be to perform the severity level processing dynamically according to monitoring objectives, such as evaluating performance or detecting anomalies.

To reach our goal, we want to deepen our understanding of logging severity to create an approach that emerges from a theory of what logging severity levels are at its core. We also hope that the results of this survey can contribute to the tasks of developers and operations engineers. The conceptual framework can be a guideline and a base to create specifications for logging practices, contributing to both groups. Moreover, an automated approach to process log data efficiently can make monitoring tasks easier for operations engineers.

\bibliographystyle{unsrtnat}
\bibliography{references}  %%% Uncomment this line and comment out the ``thebibliography'' section below to use the external .bib file (using bibtex) .

\begin{thebibliography}{31}
\providecommand{\natexlab}[1]{#1}
\providecommand{\url}[1]{\texttt{#1}}
\expandafter\ifx\csname urlstyle\endcsname\relax
  \providecommand{\doi}[1]{doi: #1}\else
  \providecommand{\doi}{doi: \begingroup \urlstyle{rm}\Url}\fi

\bibitem[El-Masri et~al.(2020)El-Masri, Petrillo, Gu{\'e}h{\'e}neuc,
  Hamou-Lhadj, and Bouziane]{IST/EL2020/systematic}
Diana El-Masri, Fabio Petrillo, Yann-Ga{\"e}l Gu{\'e}h{\'e}neuc, Abdelwahab
  Hamou-Lhadj, and Anas Bouziane.
\newblock A systematic literature review on automated log abstraction
  techniques.
\newblock \emph{Information and Software Technology}, 122:\penalty0 106276,
  2020.

\bibitem[Bushong et~al.(2020)Bushong, Sanders, Curtis, Du, Cerny, Frajtak,
  Bures, Tisnovsky, and Shin]{ICRACS/BUSHONG-2020/matching}
Vincent Bushong, Russell Sanders, Jacob Curtis, Mark Du, Tomas Cerny, Karel
  Frajtak, Miroslav Bures, Pavel Tisnovsky, and Dongwan Shin.
\newblock On matching log analysis to source code: A systematic mapping study.
\newblock In \emph{Proceedings of the International Conference on Research in
  Adaptive and Convergent Systems}, pages 181--187, 2020.

\bibitem[Yuan et~al.(2012{\natexlab{a}})Yuan, Zheng, Park, Zhou, and
  Savage]{TOCS/YUAN-2012/diagnosability}
Ding Yuan, Jing Zheng, Soyeon Park, Yuanyuan Zhou, and Stefan Savage.
\newblock Improving software diagnosability via log enhancement.
\newblock \emph{ACM Transactions on Computer Systems (TOCS)}, 30\penalty0
  (1):\penalty0 1--28, 2012{\natexlab{a}}.

\bibitem[Yao et~al.(2020)Yao, Li, Shang, and Hassan]{EMSE/YAO-2020/compressors}
Kundi Yao, Heng Li, Weiyi Shang, and Ahmed~E Hassan.
\newblock A study of the performance of general compressors on log files.
\newblock \emph{Empirical Software Engineering}, 25\penalty0 (5):\penalty0
  3043--3085, 2020.

\bibitem[Lin et~al.(2016)Lin, Zhang, Lou, Zhang, and
  Chen]{ICSE/LIN2016/log-clustering}
Qingwei Lin, Hongyu Zhang, Jian-Guang Lou, Yu~Zhang, and Xuewei Chen.
\newblock Log clustering based problem identification for online service
  systems.
\newblock In \emph{2016 IEEE/ACM 38th International Conference on Software
  Engineering Companion (ICSE-C)}, pages 102--111. IEEE, 2016.

\bibitem[Kim et~al.(2020)Kim, Kim, Park, and Park]{SPE/KIM2020/automatic}
Taeyoung Kim, Suntae Kim, Sooyong Park, and YoungBeom Park.
\newblock Automatic recommendation to appropriate log levels.
\newblock \emph{Software: Practice and Experience}, 50\penalty0 (3):\penalty0
  189--209, 2020.

\bibitem[Chen and Jiang(2017)]{ICSE/CHEN2017/characterizing-antipatterns}
Boyuan Chen and Zhen~Ming Jiang.
\newblock Characterizing and detecting anti-patterns in the logging code.
\newblock In \emph{2017 IEEE/ACM 39th International Conference on Software
  Engineering (ICSE)}, pages 71--81. IEEE, 2017.

\bibitem[Li et~al.(2018)Li, Chen, Shang, and Hassan]{EMSE-LI2018-studying}
Heng Li, Tse-Hsun~Peter Chen, Weiyi Shang, and Ahmed~E Hassan.
\newblock Studying software logging using topic models.
\newblock \emph{Empirical Software Engineering}, 23\penalty0 (5):\penalty0
  2655--2694, 2018.

\bibitem[C{\^a}ndido et~al.(2021)C{\^a}ndido, Aniche, and van
  Deursen]{PEERJ/CANDIDO-2021/LogBasedMonitoringMapping}
Jeanderson C{\^a}ndido, Maur{\'\i}cio Aniche, and Arie van Deursen.
\newblock Log-based software monitoring: a systematic mapping study.
\newblock \emph{PeerJ Computer Science}, 7:\penalty0 e489, 2021.

\bibitem[Chowdhury et~al.(2018)Chowdhury, Di~Nardo, Hindle, and
  Jiang]{ESE/CHOWDHURY2018/exploratory}
Shaiful Chowdhury, Silvia Di~Nardo, Abram Hindle, and Zhen Ming~Jack Jiang.
\newblock An exploratory study on assessing the energy impact of logging on
  android applications.
\newblock \emph{Empirical Software Engineering}, 23\penalty0 (3):\penalty0
  1422--1456, 2018.

\bibitem[Zeng et~al.(2019)Zeng, Chen, Shang, and Chen]{ESE/ZENG2019/studying}
Yi~Zeng, Jinfu Chen, Weiyi Shang, and Tse-Hsun~Peter Chen.
\newblock Studying the characteristics of logging practices in mobile apps: a
  case study on f-droid.
\newblock \emph{Empirical Software Engineering}, 24\penalty0 (6):\penalty0
  3394--3434, 2019.

\bibitem[Kabinna et~al.(2018)Kabinna, Bezemer, Shang, Syer, and
  Hassan]{ESE/KABINNA2018/examining}
Suhas Kabinna, Cor-Paul Bezemer, Weiyi Shang, Mark~D Syer, and Ahmed~E Hassan.
\newblock Examining the stability of logging statements.
\newblock \emph{Empirical Software Engineering}, 23\penalty0 (1):\penalty0
  290--333, 2018.

\bibitem[Yuan et~al.(2012{\natexlab{b}})Yuan, Park, and
  Zhou]{ICSE/YUAN2012/characterizingLoggingPractices}
Ding Yuan, Soyeon Park, and Yuanyuan Zhou.
\newblock Characterizing logging practices in open-source software.
\newblock In \emph{2012 34th International Conference on Software Engineering
  (ICSE)}, pages 102--112. IEEE, 2012{\natexlab{b}}.

\bibitem[Zhao et~al.(2017)Zhao, Rodrigues, Luo, Stumm, Yuan, and
  Zhou]{SPSP/ZHAO2017/log20}
Xu~Zhao, Kirk Rodrigues, Yu~Luo, Michael Stumm, Ding Yuan, and Yuanyuan Zhou.
\newblock Log20: Fully automated optimal placement of log printing statements
  under specified overhead threshold.
\newblock In \emph{Proceedings of the 26th Symposium on Operating Systems
  Principles}, pages 565--581, 2017.

\bibitem[Hassani et~al.(2018)Hassani, Shang, Shihab, and
  Tsantalis]{ESE/HASSANI2018/studying}
Mehran Hassani, Weiyi Shang, Emad Shihab, and Nikolaos Tsantalis.
\newblock Studying and detecting log-related issues.
\newblock \emph{Empirical Software Engineering}, 23\penalty0 (6):\penalty0
  3248--3280, 2018.

\bibitem[Li et~al.(2017)Li, Shang, and Hassan]{ESE/LI2017/LogLevelChoose}
Heng Li, Weiyi Shang, and Ahmed~E Hassan.
\newblock Which log level should developers choose for a new logging statement?
\newblock \emph{Empirical Software Engineering}, 22\penalty0 (4):\penalty0
  1684--1716, 2017.

\bibitem[He et~al.(2018)He, Chen, He, and
  Lyu]{ASE/HE2018/characterizingNaturalLanguageDescriptions}
Pinjia He, Zhuangbin Chen, Shilin He, and Michael~R Lyu.
\newblock Characterizing the natural language descriptions in software logging
  statements.
\newblock In \emph{2018 33rd IEEE/ACM International Conference on Automated
  Software Engineering (ASE)}, pages 178--189. IEEE, 2018.

\bibitem[Yuan et~al.(2019)Yuan, Shi, Liang, and Qin]{CLOUD/YUAN2019/approach}
Yue Yuan, Wenchang Shi, Bin Liang, and Bo~Qin.
\newblock An approach to cloud execution failure diagnosis based on exception
  logs in openstack.
\newblock In \emph{2019 IEEE 12th International Conference on Cloud Computing
  (CLOUD)}, pages 124--131. IEEE, 2019.

\bibitem[Gholamian and Ward(2021)]{ARXIV/GHOLAMIAN-2021/comprehensive}
Sina Gholamian and Paul~AS Ward.
\newblock A comprehensive survey of logging in software: From logging
  statements automation to log mining and analysis.
\newblock \emph{arXiv preprint arXiv:2110.12489}, 2021.

\bibitem[Mendes and Petrillo(2021)]{ARXIV/MENDES-2021/towards-noisiness}
Eduardo Mendes and Fabio Petrillo.
\newblock Towards logging noisiness theory: quality aspects to characterize
  unwanted log entries.
\newblock \emph{arXiv preprint arXiv:2106.03018}, 2021.

\bibitem[Rong et~al.(2018)Rong, Gu, Zhang, Shao, and
  Liu]{ASWEC/RONG2018/logging}
Guoping Rong, Shenghui Gu, He~Zhang, Dong Shao, and Wanggen Liu.
\newblock How is logging practice implemented in open source software projects?
  a preliminary exploration.
\newblock In \emph{2018 25th Australasian Software Engineering Conference
  (ASWEC)}, pages 171--180. IEEE, 2018.

\bibitem[Ding et~al.(2015)Ding, Zhou, Lou, Zhang, Lin, Fu, Zhang, and
  Xie]{USENIX/DING-2015/log2}
Rui Ding, Hucheng Zhou, Jian-Guang Lou, Hongyu Zhang, Qingwei Lin, Qiang Fu,
  Dongmei Zhang, and Tao Xie.
\newblock Log2: A cost-aware logging mechanism for performance diagnosis.
\newblock In \emph{2015 $\{$USENIX$\}$ Annual Technical Conference
  ($\{$USENIX$\}$$\{$ATC$\}$ 15)}, pages 139--150, 2015.

\bibitem[Oliner et~al.(2012)Oliner, Ganapathi, and
  Xu]{COMACM/OLINER2012/advances}
Adam Oliner, Archana Ganapathi, and Wei Xu.
\newblock Advances and challenges in log analysis.
\newblock \emph{Communications of the ACM}, 55\penalty0 (2):\penalty0 55--61,
  2012.

\bibitem[Anu et~al.(2019)Anu, Chen, Shi, Hou, Liang, and
  Qin]{ICSME/ANU2019/verbosityloglevels}
Han Anu, Jie Chen, Wenchang Shi, Jianwei Hou, Bin Liang, and Bo~Qin.
\newblock An approach to recommendation of verbosity log levels based on
  logging intention.
\newblock In \emph{2019 IEEE International Conference on Software Maintenance
  and Evolution (ICSME)}, pages 125--134. IEEE, 2019.

\bibitem[Li et~al.(2020{\natexlab{a}})Li, Chen, and
  Shang]{ACMCASE/LI2020/where-shall}
Zhenhao Li, Tse-Hsun Chen, and Weiyi Shang.
\newblock Where shall we log? studying and suggesting logging locations in code
  blocks.
\newblock In \emph{Proceedings of the 35th IEEE/ACM International Conference on
  Automated Software Engineering}, pages 361--372, 2020{\natexlab{a}}.

\bibitem[Fu et~al.(2014)Fu, Zhu, Hu, Lou, Ding, Lin, Zhang, and
  Xie]{ICSE/FU2014/developers}
Qiang Fu, Jieming Zhu, Wenlu Hu, Jian-Guang Lou, Rui Ding, Qingwei Lin, Dongmei
  Zhang, and Tao Xie.
\newblock Where do developers log? an empirical study on logging practices in
  industry.
\newblock In \emph{Companion Proceedings of the 36th International Conference
  on Software Engineering}, pages 24--33, 2014.

\bibitem[Li et~al.(2020{\natexlab{b}})Li, Shang, Adams, Sayagh, and
  Hassan]{TSE/LI2020/qualitative}
Heng Li, Weiyi Shang, Bram Adams, Mohammed Sayagh, and Ahmed~E Hassan.
\newblock A qualitative study of the benefits and costs of logging from
  developers' perspectives.
\newblock \emph{IEEE Transactions on Software Engineering}, 2020{\natexlab{b}}.

\bibitem[Zhu et~al.(2015)Zhu, He, Fu, Zhang, Lyu, and
  Zhang]{ICSE/ZHU-2015/learning}
Jieming Zhu, Pinjia He, Qiang Fu, Hongyu Zhang, Michael~R Lyu, and Dongmei
  Zhang.
\newblock Learning to log: Helping developers make informed logging decisions.
\newblock In \emph{2015 IEEE/ACM 37th IEEE International Conference on Software
  Engineering}, volume~1, pages 415--425. IEEE, 2015.

\bibitem[Gholamian and Ward(2020)]{ACMSAC/GHOLAMIAN2020/logging}
Sina Gholamian and Paul~AS Ward.
\newblock Logging statements' prediction based on source code clones.
\newblock In \emph{Proceedings of the 35th Annual ACM Symposium on Applied
  Computing}, pages 82--91, 2020.

\bibitem[Jia et~al.(2018)Jia, Li, Liu, Liao, and Liu]{ICSA/JIA-2018/smartlog}
Zhouyang Jia, Shanshan Li, Xiaodong Liu, Xiangke Liao, and Yunhuai Liu.
\newblock Smartlog: Place error log statement by deep understanding of log
  intention.
\newblock In \emph{2018 IEEE 25th International Conference on Software
  Analysis, Evolution and Reengineering (SANER)}, pages 61--71. IEEE, 2018.

\bibitem[Roudjane et~al.(2021)Roudjane, Reba{\"\i}ne, Khoury, and
  Hall{\'e}]{IS/ROUDJANE-2021/detecting}
Massiva Roudjane, Djamal Reba{\"\i}ne, Rapha{\"e}l Khoury, and Sylvain
  Hall{\'e}.
\newblock Detecting trend deviations with generic stream processing patterns.
\newblock \emph{Information Systems}, 101:\penalty0 101446, 2021.

\end{thebibliography}





% \end{thebibliography}

\end{document}